\documentclass[aps,prd,twocolumn,showpacs,groupedaddress]{revtex4}  

\usepackage{graphicx}
\usepackage{booktabs}
\usepackage{amssymb,bm,mathrsfs,bbm,amscd}
\usepackage[tbtags]{amsmath}

\begin{document}

\preprint{Submitted to PRD}

\title{Enhanced sensitivity to Lorentz invariance violations in short-range gravity experiments}
\author{Cheng-Gang Shao}
\author{ Ya-Fen Chen}
\author{Yu-Jie Tan}
\author{Jun Luo}
\author{Shan-Qing Yang}\email[E-mail:]{ysq2011@hust.edu}

\affiliation
{Key Laboratory of Fundamental Physical Quantities
Measurement of Ministry of Education, School of Physics,
 Huazhong University of Science and Technology, Wuhan 430074, People's
Republic of China}

\author{Michael Edmund Tobar} \email[E-mail:]{michael.tobar@uwa.edu.au}

\affiliation
{School of Physics, University of Western Australia, Crawley, WA 6009, Australia}

\date{today}

\begin{abstract}
Recently, first limits on putative Lorentz invariance violation coefficients in the pure gravity sector were determined by the reanalysis of short-range gravity experiments. Such experiments search for new physics at sidereal frequencies. They are not, however, designed to optimize the signal strength of a Lorentz invariance violation force; in fact the Lorentz violating signal is suppressed in the planar test mass geometry employed in those experiments. We describe a short-range torsion pendulum experiment with enhanced sensitivity to possible Lorentz violating signals. A periodic, striped test mass geometry is used to augment the signal. Careful arrangement of the phases of the striped patterns on opposite ends of the pendulum further enhances the signal while simultaneously suppressing the Newtonian background.
\end{abstract}


\maketitle
\section{Introduction}

Local Lorentz symmetry is a solid foundation of General Relativity (GR), which is a classical theory and may demand some changes in its foundational structure to merge gravity with quantum mechanics. There are many theoretical scenarios in which this symmetry might be broken. Even if local Lorentz invariance is exact in the underlying theory of quantum gravity, spontaneous breaking of this symmetry may occur, leading to miniscule observable effects \cite{1,2}. Additionally, Lorentz invariance violations (LIV) may be relatively large but hidden from the many experiments to date \cite{3,4}, which  set very tight bounds across many physical sectors (including the photon, proton, electron and other particle sectors). Thus, in general, the investigation of LIV is a valuable tool to probe the foundations of GR \cite{5,6}  without preconceived notions of the numerical sensitivity.

A more recent area for searching for LIV is in the pure gravity sector \cite{7}, with some recent first bounds utilizing either short-range torsional experiments \cite{8,9,9a}, planetary motion \cite{9b}, gravitational Cerenkov radiation \cite{9c} and gravitational wave dispersion \cite{9d}. Short-range experiments at Indiana University (IU) \cite{8}  and Huazhong University of Science and Technology (HUST) \cite{9e} have achieved sensitivities at the level of $10^{-7}$ to $10^{-8}$ $\rm{ m^2}$ respectively, to individual coefficients controlling these types of gravitational LIV. More recently a combined analysis from the same two experimental groups permitted the first simultaneous limits on the 14 nonrelativistic LIV coefficients in the pure gravity sector, at the level of $10^{-9}$ $\rm{ m^2}$ \cite{9a}. However, these experiments use planar geometry, as they are designed primarily to search for Yukawa-type non-Newtonian gravity \cite{9c}. The LIV force between two parallel, coaxial plates is suppressed, and in fact vanishes in the limit that one plate has infinite area \cite{9}.

In this work, we analyse a modified short-range gravity experiment with enhanced sensitivity to LIV. The phase of the striped pattern on the pendulum test mass is shifted relative to the pattern of the immediately opposite source mass. An "asymmetric" design, in which this phase relationship is reversed for the pair of masses on the opposite end of the pendulum, further enhances the LIV signal and suppresses the Newtonian background.

\section{Lorentz invariance violation with quadratic curvature couplings}

It has been shown that in gravitation experiments, the perturbative corrections to the Newtonian potential between two test masses $m_1$  and $m_2$  in the Standard-Model Extension (SME) is given by \cite{7}
\begin{eqnarray}\label{equation1:eps}
{V_{LV}}(\vec{r})=- G\frac{{m_1}{m_2}}{{\left| \vec{r} \right|}^3}\bar k(\hat r,T).\label{Eq:1}
\end{eqnarray}
Here the vector $\vec{r}=\vec{r}_1-\vec{r}_2$ separates $m_1$ and $ m_2$, and
\begin{eqnarray}\label{equation2:eps}
\bar k(\hat r,\!T\!)\!\!=\!\!\frac{3}{2}{(\!{\bar k_{e\!f\!f}}\!)_{jkjk}}\!\! - \!\!9{(\!{\bar k_{e\!f\!f}}\!)_{jkll}}{\hat r^j}{\hat r^k}\!\!
+\!\!\! \frac{{15}}{2}{(\!{\bar k_{e\!f\!f\!}}\!)_{jklm}}{\hat r^j}{\hat r^k}{\hat r^l}{\hat r^m}.\nonumber\\
\end{eqnarray} with dimensions of squared length, containing 15 independent degrees of freedom of which 14 are measurable\cite{9a}.

The Sun-centered celestial-equatorial frame has become the convention for reporting results from experimental searches for the LIV coefficients across all SME sectors. In this frame the $Z$ axis points along the direction of the Earth's rotation axis and the $X$ axis points towards the vernal equinox \cite{10,11,12,13}. Thus, the rotation of the Earth modulates the LIV coefficients in the laboratory frame, producing putative sidereal signals in the experimental data \cite{14}. Ignoring any boost dependence (suppressed by $10^{-4}$), the conversion from the Sun-centered frame ($X, Y, Z$) to the laboratory frame ($x, y, z$) with $x$ axis pointing to local south and $z$ axis to the zenith can be implemented by the time-dependent rotation
\begin{equation}
{R^{jJ}}=\left({\begin{array}{*{20}{c}}
{\cos\chi \cos {\omega _ \oplus}T}&{\cos\chi\sin{\omega _\oplus}T}&{-\sin\chi}\\
{ -\sin {\omega _\oplus}T}&{\cos {\omega _\oplus }T}&0\\
{\sin\chi\cos{\omega _\oplus}T}&{\sin\chi\sin{\omega _\oplus}T}&{\cos\chi}
\end{array}}\right). \label{Eq:3}
\end{equation}
Here the angle $\chi$ is the colatitude of the laboratory and ${\omega _ \oplus } \simeq 2\pi /(23.93$ $ {\rm{h}})$ is the Earth's sidereal frequency. The sidereal time $T$-dependent coefficients ${({\bar k_{eff}})_{jklm}}$ are thus related to the constant coefficients ${({\bar k_{eff}})_{JKLM}}$ in the Sun-centered frame by
\begin{equation}
{({\bar k_{eff}})_{jklm}} = {R^{jJ}}{R^{kK}}{R^{lL}}{R^{mM}}{({\bar k_{eff}})_{JKLM}}.\label{Eq:4}
\end{equation}
Therefore, the inverse-cube potential for LIV is oscillatory with $T$ and includes components up to the fourth harmonic of $\omega _\oplus$ in laboratory gravity experiments. Equivalently, the LIV force between two point masses can be expected to vary with frequencies up to and including the fourth harmonic of $\omega _\oplus$. Thus, $\bar k(\hat r,T)$ can be expressed as the following Fourier series 
\begin{equation}
\bar k(\hat r,T) = {c_0} + \sum\limits_{m = 1}^4 {[{c_m}\cos (m{\omega _ \oplus }T) + {s_m}\sin (m{\omega _ \oplus }T)]} .\label{Eq:5}
\end{equation}
The nine Fourier amplitudes in this expression are functions of ${({\bar k_{eff}})_{JKLM}}$ according to Eq. (4).

In short-range experiments, one can compute the nine Fourier coefficients from time-dependent force data to search for LIV. To simplify the analysis, we separate the even and odd harmonics in Eq. 5 with the introduction of the modified coefficients $\bar {k}_j$, where $j = 0,1,2, \cdots ,14$. The $\bar {k}_j$ are 15 independent linear combinations of the ${({\bar k_{eff}})_{JKLM}}$. The first nine coefficients represent even harmonics,
{\setlength\belowdisplayskip{18pt}
\begin{equation}
\left( {\begin{array}{*{20}{c}}
{{\bar k}_0}\\
{{\bar k}_1}\\
{{\bar k}_2}\\
{{\bar k}_3}\\
{{\bar k}_4}\\
{{\bar k}_5}\\
{{\bar k}_6}\\
{{\bar k}_7}\\
{{\bar k}_8}
\end{array}}\right)\!\!\! =\!\!\!\left(\!\!\!{\begin{array}{*{20}{c}}
1&0&0&0&0&0&0&0&0\\
0&1&1&2&0&0&0&0&0\\
0&0&0&0&1&1&0&0&0\\
0&1&{-1}&0&0&0&0&0&0\\
0&0&0&0&1&{-1}&0&0&0\\
0&0&0&0&0&0&1&1&0\\
0&0&0&0&0&0&0&0&1\\
0&{-1}&{-1}&6&0&0&0&0&0\\
0&0&0&0&0&0&1&{-1}&0
\end{array}}\!\!\!\right)\!\!\!\left(\!\!{\begin{array}{*{20}{c}}
{{({{\bar k}_{eff}})}_{ZZZZ}}\\
{{({{\bar k}_{eff}})}_{XXXX}}\\
{{({{\bar k}_{eff}})}_{YYYY}}\\
{{({{\bar k}_{eff}})}_{XXYY}}\\
{{({{\bar k}_{eff}})}_{XXZZ}}\\
{{({{\bar k}_{eff}})}_{YYZZ}}\\
{{({{\bar k}_{eff}})}_{XXXY}}\\
{{({{\bar k}_{eff}})}_{XYYY}}\\
{{({{\bar k}_{eff}})}_{XYZZ}}
\end{array}}\!\! \right).\label{Eq:6}
\end{equation}
The remaining coefficients represent odd harmonics,
\begin{eqnarray}
\left( {\begin{array}{*{20}{c}}
  {{{\bar k}_9}} \\
  {{{\bar k}_{10}}} \\
  {{{\bar k}_{11}}} \\
  {{{\bar k}_{12}}} \\
  {{{\bar k}_{13}}} \\
  {{{\bar k}_{14}}}
\end{array}} \right)=\left( {\begin{array}{*{20}{c}}
  1&0&0&0&0&0 \\
  0&1&0&0&0&0 \\
  0&0&1&0&0&1 \\
  0&0&0&1&1&0\\
  0&0&1&0&0&{ - 3}\\
  0&0&0&1&{ - 3}&0
\end{array}} \right)\left( {\begin{array}{*{20}{c}}
  {{{({{\bar k}_{eff}})}_{XZZZ}}} \\
  {{{({{\bar k}_{eff}})}_{YZZZ}}} \\
  {{{({{\bar k}_{eff}})}_{XXXZ}}} \\
  {{{({{\bar k}_{eff}})}_{YYYZ}}} \\
  {{{({{\bar k}_{eff}})}_{XXYZ}}} \\
  {{{({{\bar k}_{eff}})}_{XYYZ}}}
\end{array}} \right).\label{Eq:7}
\end{eqnarray}
Combining the transformation matrix $R^{jJ}$ and Eqs. (6) and (7), the nine Fourier amplitudes in Eq. (5) take the explicit form;
{\setlength\abovedisplayskip{15pt}
{\setlength\belowdisplayskip{15pt}
\begin{eqnarray}\label{equation8:eps}
\begin{array}{l}
{c_0} = {\alpha _0}{\bar k_0} + {\alpha _1}{\bar k_1} + {\alpha _2}{\bar k_2}\\[2mm]
{c_2} = {\alpha _3}{\bar k_3} + {\alpha _4}{\bar k_4} + {\alpha _5}{\bar k_5} + {\alpha _6}{\bar k_6}\\[2mm]
{s_2} =  - \frac{1}{2}{\alpha _5}{\bar k_3} - \frac{1}{2}{\alpha _6}{\bar k_4} + 2{\alpha _3}{\bar k_5} + 2{\alpha _4}{\bar k_6}\\[2mm]
{c_4} = {\alpha _7}{\bar k_7} + {\alpha _8}{\bar k_8}\\[2mm]
{s_4} = \frac{1}{4}{\alpha _8}{\bar k_7} - 4{\alpha _7}{\bar k_8}
\end{array}
\end{eqnarray}
for even harmonics, involving 9 functions ${\alpha _j}(\hat r,\chi )$ with $j = 0,1,2, \cdots ,8$. and
{\setlength\abovedisplayskip{15pt}
{\setlength\belowdisplayskip{15pt}
\begin{eqnarray}\label{equation9:eps}
\begin{array}{l}
{c_1} = {\alpha _9}{\bar k_9} + {\alpha _{10}}{\bar k_{10}} + {\alpha _{11}}{\bar k_{11}} + {\alpha _{12}}{\bar k_{12}}\\[2mm]
{s_1} =  - {\alpha _{10}}{\bar k_9} + {\alpha _9}{\bar k_{10}} - {\alpha _{12}}{\bar k_{11}} + {\alpha _{11}}{\bar k_{12}}\\[2mm]
{c_3} = {\alpha _{13}}{\bar k_{13}} + {\alpha _{14}}{\bar k_{14}}\\[2mm]
{s_3} = {\alpha _{14}}{\bar k_{13}} - {\alpha _{13}}{\bar k_{14}}
\end{array}
\end{eqnarray}
for odd harmonics, involving 6 functions
${\alpha _j}(\hat r,\chi )$ with $j = 9,10, \cdots ,14$. The Fourier amplitude $c_0$ is a linear combination of ${\bar{ k}_0}$, ${\bar{ k}_1}$, and ${\bar {k}_2}$. The amplitudes $ c_2$ and $s_2$ are linear combinations of ${\bar {k}_3}$, ${\bar {k}_4}$, ${\bar{ k}_5}$ and ${\bar {k}_6}$. Amplitudes $c_4$ and $s_4$ are the linear combinations of ${\bar {k}_7}$ and ${\bar{ k}_8}$. Similar relations apply to the odd harmonics.

By introducing the rotation
\begin{equation}\label{Eq:10}
\left\{\begin{array}{ll}
\tilde x = x\cos \chi  + z\sin \chi \\[2mm]
\tilde z =  - x\sin \chi  + z\cos \chi
\end{array}\right.
\end{equation}
the corresponding expressions ${\alpha _j}(\hat r,\chi )$, can be written as
\begin{equation} \label{equation11:eps}
\begin{array}{l}
{\alpha _0} (\hat r,\chi )= \displaystyle\frac{3}{2} - \displaystyle\frac{9}{{{r^2}}}{\tilde z^2}+ \displaystyle\frac{{15}}{{2{r^4}}}{\tilde z ^4}\\[2mm]
{\alpha_1}(\hat r,\chi )=-\displaystyle\frac{3}{{16}}-\displaystyle\frac{9}{{8{r^2}}}{\tilde z^2}+\displaystyle\frac{{45}}{{16{r^4}}}{\tilde z^4}\\[2mm]
{\alpha_2}(\hat r,\chi )=-\displaystyle\frac{3}{2}+\displaystyle\frac{{18}}{{{r^2}}}{\tilde z^2}-\displaystyle\frac{{45}}{{2{r^4}}}{\tilde z^4}\\[2mm]
{\alpha_3}(\hat r,\chi )=-\displaystyle\frac{9}{2}\displaystyle\frac{{{{\tilde x}^2}-{y^2}}}{{{r^2}}}+\displaystyle\frac{{15}}{4}\displaystyle\frac{{{{\tilde x}^4}-{y^4}}}{{{r^4}}}\\[2mm]
{\alpha _4} (\hat r,\chi )=  - \displaystyle\frac{9}{2}\displaystyle\frac{{{{\tilde x}^2} - {y^2}}}{{{r^2}}}\left( {1 - 5\displaystyle\frac{{{{\tilde x}^2}}}{{{r^2}}}} \right)\\[2mm]
{\alpha_5}(\hat r,\chi )=\displaystyle\frac{{\tilde xy}}{{{r^2}}}\left({-18+15\displaystyle\frac{{{{\tilde x}^2}+{y^2}}}{{{r^2}}}}\right)\\[2mm]
{\alpha _6}(\hat r,\chi ) = 18\displaystyle\frac{{\tilde xy}}{{{r^2}}}\left( { - 1 + 5\displaystyle\frac{{{{\tilde z}^2}}}{{{r^2}}}} \right)\\[2mm]
{\alpha _7}(\hat r,\chi ) = \displaystyle\frac{{45}}{8}\frac{{{{\tilde x}^2}{y^2}}}{{{r^4}}} - \displaystyle\frac{{15}}{{16}}\displaystyle\frac{{{{\tilde x}^4} + {y^4}}}{{{r^4}}}\\[2mm]
{\alpha _8}(\hat r,\chi ) = 15\displaystyle\frac{{\tilde xy}}{{{r^2}}}\displaystyle\frac{{{{\tilde x}^2} - {y^2}}}{{{r^2}}}\\[2mm]
{\alpha _9}(\hat r,\chi ) = \displaystyle\frac{{-\tilde x\tilde z}}{{{r^2}}}( {18 - 30\displaystyle\frac{{{{\tilde z}^2}}}{{{r^2}}}})\\[2mm]
{\alpha _{10}}(\hat r,\chi ) =  \displaystyle\frac{{\tilde z y}}{{{r^2}}}\left( { - 18 + 30\displaystyle\frac{{{{\tilde z }^2}}}{{{r^2}}}} \right)\\[2mm]
{\alpha _{11}}(\hat r,\chi ) = \displaystyle\frac{{-\tilde x\tilde z}}{{{r^2}}}\left( {18 - \displaystyle\frac{{45}}{2}\displaystyle\frac{{{{\tilde x}^2} + {y^2}}}{{{r^2}}}}\right)\\[2mm]
{\alpha _{12}}(\hat r,\chi ) = \displaystyle\frac{{-\tilde zy}}{{{r^2}}}\left( {18 - \displaystyle\frac{{45}}{2}\displaystyle\frac{{{{\tilde x}^2} + {y^2}}}{{{r^2}}}} \right)\\[2mm]
{\alpha _{13}}(\hat r,\chi ) = \displaystyle\frac{{15}}{2}\displaystyle\frac{{-\tilde x\tilde z}}{{{r^2}}}\displaystyle\frac{{3{y^2}-{{\tilde x}^2}}}{{{r^2}}}\\[2mm]
{\alpha_{14}}(\hat r,\chi )=\displaystyle\frac{{45}}{2}\displaystyle\frac{{-\tilde zy}}{{{r^2}}}\displaystyle\frac{{{{\tilde x}^2}-{y^2}/3}}{{{r^2}}}
\end{array}
\end{equation}
From Eq. (11), we obtain
\begin{equation}
3{\alpha _0} + 8{\alpha _1} + 2{\alpha _2} = 0,\label{Eq:12}
\end{equation}
and the constant term $c_0$ can be expressed
\begin{equation}
{c_0} = {\alpha _1}({\bar k_1} - \frac{8}{3}{\bar k_0}) + {\alpha _2}({\bar k_2} - \frac{2}{3}{\bar k_0}). \label{Eq:13}
\end{equation}
This means that the nine Fourier amplitudes $c_i$ and $ s_i$ can be expressed as linear combinations of the $({\bar k_{eff}})_{JKLM}$ via the reduced set of 14 independent functions ${\alpha _j}(\hat r,\chi )$ with $j = 1,2, \cdots ,14$. In other words, only 14 degrees of freedom of $({\bar k_{eff}})_{JKLM}$ are independently measurable in short-range experiments. The remaining one is the double trace $({\bar k_{eff}})_{JKJK}$, which is rotation invariant, and produces only a contact correction to the usual Newtonian force.
In summary, we have introduced the ${\bar{ k}_{j}}$ to decompose the 14 dimensional $({\bar k_{eff}})_{JKLM}$ into 5 subspaces (Eq. 8). The main advantage is that it isolates the different harmonics of the LIV, making unique harmonic violation signals correspond to different subspaces.\\

\section{Measurable coefficients in Lorentz invariance violation}

The double trace ${({\bar k_{e\!f\!f}})_{JKJK}}$ is a rotational scalar, here denoted by $u$. After separating this scalar degree of freedom, we decompose the coefficients ${({\bar k_{e\!f\!f}})_{JKLM}}$ for the LIV according to
\begin{eqnarray}\label{equation14:eps}
{(\bar k_{e\!f\!f})_{J\!K\!L\!M}\!\!=\!\!{({\tilde{k}_{e\!f\!f}})_{J\!K\!L\!M}}\!\!
+\!\!\frac{u}{15}({{{\delta\! _{J\!K}}{\delta\! _{L\!M}}\!\! +\! \!{\delta\! _{J\!L}}{\delta _{K\!M}}\!\! +\!\! {\delta\! _{J\!M}}{\delta \!_{K\!L}}}})},\nonumber\\
\end{eqnarray}
with ${({\tilde k_{e\!f\!f}})_{J\!K\!J\!K}}\!\!\! =\!\!0$. It is straightforward to show that the scalar part, also denoted
$\frac{1}{15}u({{{\delta _{jk}}{\delta _{lm}} + {\delta _{jl}}{\delta _{km}} + {\delta _{jm}}{\delta _{kl}}}})$ in the laboratory frame, has no contribution to the $\bar{ k}(\hat r,T)$ in Eq.(2). Therefore, the 14 independently measurable coefficients are $({\tilde k_{e\!f\!f}})_{JKLM}$. Equivalently, we may adopt
${({\bar k_{e\!f\!f}})_{JKLM}} - \frac{1}{3}{({\bar k_{e\!f\!f}})_{ZZZZ}}({{{\delta _{JK}}{\delta _{LM}} + {\delta _{JL}}{\delta _{KM}} + {\delta _{JM}}{\delta _{KL}}}})$ for the 14 measurable coefficients in Sun-centered framed, since the scalar part is not observable in Eq. (2).

Combining Eqs. (13) and (6), we redefine the eight ${\bar{ k}_j}$ with $j = 1,2, \cdots ,8$ as
\begin{eqnarray}
\left( {\begin{array}{*{20}{c}}
  {{{\bar k}_1}} \\
  {{{\bar k}_2}} \\
  {{{\bar k}_3}} \\
  {{{\bar k}_4}} \\
  {{{\bar k}_5}} \\
  {{{\bar k}_6}} \\
  {{{\bar k}_7}} \\
  {{{\bar k}_8}}
\end{array}} \right) = R_{even}
\left( {\begin{array}{*{20}{c}}
  {{{({{\bar k}_{eff}})}_{XXXX}} - {{({{\bar k}_{eff}})}_{ZZZZ}}} \\
  {{{({{\bar k}_{eff}})}_{YYYY}} - {{({{\bar k}_{eff}})}_{ZZZZ}}} \\
  {{{({{\bar k}_{eff}})}_{XXYY}} - \frac{1}{3}{{({{\bar k}_{eff}})}_{ZZZZ}}} \\
  {{{({{\bar k}_{eff}})}_{XXZZ}} - \frac{1}{3}{{({{\bar k}_{eff}})}_{ZZZZ}}} \\
  {{{({{\bar k}_{eff}})}_{YYZZ}} - \frac{1}{3}{{({{\bar k}_{eff}})}_{ZZZZ}}} \\
  {{{({{\bar k}_{eff}})}_{XXXY}}} \\
  {{{({{\bar k}_{eff}})}_{XYYY}}} \\
  {{{({{\bar k}_{eff}})}_{XYZZ}}}
\end{array}} \right) \label{Eq:15}
\end{eqnarray}
with matrix
\begin{eqnarray}
 R_{even} = \left( {\begin{array}{*{20}{c}}
  1&1&2&0&0&0&0&0 \\
  0&0&0&1&1&0&0&0 \\
  1&{ - 1}&0&0&0&0&0&0 \\
  0&0&0&1&{ - 1}&0&0&0 \\
  0&0&0&0&0&1&1&0 \\
  0&0&0&0&0&0&0&1 \\
  { - 1}&{ - 1}&6&0&0&0&0&0 \\
  0&0&0&0&0&1&{ - 1}&0
\end{array}} \right) \label{Eq:16}
\end{eqnarray}
for even harmonics. Inserting Eqs. (15) and (16) into Eq. (8), five Fourier amplitudes can be expressed as
\begin{widetext}
\begin{eqnarray} \label{equation17:eps}
\left( {\begin{array}{*{20}{c}}
  {{c_0}} \\
  {{c_2}} \\
  {{s_2}} \\
  {{c_4}} \\
  {{s_4}}
\end{array}} \right) \!= \left( {\begin{array}{*{20}{c}}
  {{\alpha _1}}&{{\alpha _1}}&{2{\alpha _1}}&{{\alpha _2}}&{{\alpha _2}}&0&0&0 \\
  {{\alpha _3}}&{ - {\alpha _3}}&0&{{\alpha _4}}&{ - {\alpha _4}}&{{\alpha _5}}&{{\alpha _5}}&{{\alpha _6}} \\
  { - \frac{1}{2}{\alpha _5}}&{\frac{1}{2}{\alpha _5}}&0&{ - \frac{1}{2}{\alpha _6}}&{\frac{1}{2}{\alpha _6}}&{2{\alpha _3}}&{2{\alpha _3}}&{2{\alpha _4}} \\
  { - {\alpha _7}}&{ - {\alpha _7}}&{6{\alpha _7}}&0&0&{{\alpha _8}}&{ - {\alpha _8}}&0 \\
  { - \frac{1}{4}{\alpha _8}}&{ - \frac{1}{4}{\alpha _8}}&{\frac{3}{2}{\alpha _8}}&0&0&{ - 4{\alpha _7}}&{4{\alpha _7}}&0
\end{array}} \right)\left( {\begin{array}{*{20}{c}}
  {{{({{\bar k}_{eff}})}_{XXXX}} - {{({{\bar k}_{eff}})}_{ZZZZ}}} \\
  {{{({{\bar k}_{eff}})}_{YYYY}} - {{({{\bar k}_{eff}})}_{ZZZZ}}} \\
  {{{({{\bar k}_{eff}})}_{XXYY}} - \frac{1}{3}{{({{\bar k}_{eff}})}_{ZZZZ}}} \\
  {{{({{\bar k}_{eff}})}_{XXZZ}} - \frac{1}{3}{{({{\bar k}_{eff}})}_{ZZZZ}}} \\
  {{{({{\bar k}_{eff}})}_{YYZZ}} - \frac{1}{3}{{({{\bar k}_{eff}})}_{ZZZZ}}} \\
  {{{({{\bar k}_{eff}})}_{XXXY}}} \\
  {{{({{\bar k}_{eff}})}_{XYYY}}} \\
  {{{({{\bar k}_{eff}})}_{XYZZ}}}
\end{array}} \right).
\end{eqnarray}
\end{widetext}
The six ${\bar{ k}_j}$ with $j = 9,10, \cdots ,14$ defined in Eq. (7), need not be changed. According to Eq. (9), the Fourier amplitudes for the odd harmonics are
\begin{eqnarray}
 &&\left(\!\!\! {\begin{array}{*{20}{c}}
  {{c_1}} \\
  {{s_1}} \\
  {{c_3}} \\
  {{s_3}}
\end{array}}\!\!\! \right)\!\!\! =\!\!\! \left(\!\!\!\! {\begin{array}{*{10}{c}}
  {{\alpha _9}}\!\!\! &{{\alpha _{10}}}\!\! \!&{{\alpha _{11}}}\!\! \!&{{\alpha _{12}}}\!\!\! &{{\alpha _{12}}}\!\!\! &{{\alpha _{11}}} \\
  { - {\alpha _{10}}}\!\!\! &{{\alpha _9}}\!\!\! &{ - {\alpha _{12}}}\!\!\! &{{\alpha _{11}}}\!\!\! &{{\alpha _{11}}}\!\!\! &{ - {\alpha _{12}}} \\
  0&\!\!\! 0\!\!\! &{{\alpha _{13}}}\!\!\! &{{\alpha _{14}}}\!\!\! &{ - 3{\alpha _{14}}}\!\!\! &{ - 3{\alpha _{13}}} \\
  0&\!\!\! 0\!\!\! &{{\alpha _{14}}}\!\!\! &{ - {\alpha _{13}}}\!\!\! &{3{\alpha _{13}}}\!\!\! &{ - 3{\alpha _{14}}}
\end{array}}\!\!\! \right)\!\!\!\!
\left(\!\!\! {\begin{array}{*{20}{c}}
  {{{({{\bar k}_{e\!f\!f}})}_{X\!Z\!Z\!Z}}} \\
  {{{({{\bar k}_{e\!f\!f}})}_{Y\!Z\!Z\!Z}}} \\
  {{{({{\bar k}_{e\!f\!f}})}_{X\!X\!X\!Z}}} \\
  {{{({{\bar k}_{e\!f\!f}})}_{Y\!Y\!Y\!Z}}} \\
  {{{({{\bar k}_{e\!f\!f}})}_{X\!X\!Y\!Z}}} \\
  {{{({{\bar k}_{e\!f\!f}})}_{X\!Y\!Y\!Z}}}
\end{array}} \!\!\!\right) \nonumber
\\
\end{eqnarray}

We note that if we adopt ${({\tilde k_{eff}})_{JKLM}}$ as the 14 measurable coefficients, we can introduce coefficients ${\tilde{ k}_j}$ with $j = 0,1,2, \cdots ,14$ similar to Eqs. (6) and (7), where $\bar {k}_j$ and ${({\bar k_{eff}})_{JKLM}}$ are by $\tilde {k}_j$ and $({\tilde k_{eff}})_{JKLM}$, respectively. Inserting the double trace condition
\begin{equation}
{({\tilde k_{eff}})_{JKJK}} = {\tilde k_0} + {\tilde k_1} + 2{\tilde k_2} = 0\label{Eq:19}
\end{equation}
into Eq. (13), the constant term can be further written as
\begin{eqnarray}\label{equation20:eps}
{c_0} = {\alpha _1}({\tilde k_1} - \frac{8}{3}{\tilde k_0}) + {\alpha _2}({\tilde k_2} - \frac{2}{3}{\tilde k_0})= {\beta _1}{\tilde k_1} + {\beta _2}{\tilde k_2},
\end{eqnarray}
where we have introduced
\begin{equation}\label{Eq:21}
\left\{\begin{array}{cc}
{\beta _1} = \frac{{11}}{3}{\alpha _1} + \frac{2}{3}{\alpha _2} \\[2mm]
{\beta _2} = \frac{{16}}{3}{\alpha _1} + \frac{7}{3}{\alpha _2}.
\end{array}\right.
\end{equation}
The relation between the nine Fourier amplitudes and $(\tilde k_{eff})_{JKLM}$ is similar to Eqs. (17) and (18), with ${\alpha _1}$ and ${\alpha _2}$ being replaced by ${\beta _1}$ and ${\beta _2}$, respectively. The explicit expressions for odd and even harmonics are
\begin{eqnarray}
 &&\left(\!\!\! {\begin{array}{*{20}{c}}
  {{c_1}} \\
  {{s_1}} \\
  {{c_3}} \\
  {{s_3}}
\end{array}}\!\!\! \right)\!\!\! =\!\!\! \left(\!\!\!\! {\begin{array}{*{10}{c}}
  {{\alpha _9}}\!\!\! &{{\alpha _{10}}}\!\! \!&{{\alpha _{11}}}\!\! \!&{{\alpha _{12}}}\!\!\! &{{\alpha _{12}}}\!\!\! &{{\alpha _{11}}} \\
  { - {\alpha _{10}}}\!\!\! &{{\alpha _9}}\!\!\! &{ - {\alpha _{12}}}\!\!\! &{{\alpha _{11}}}\!\!\! &{{\alpha _{11}}}\!\!\! &{ - {\alpha _{12}}} \\
  0&\!\!\! 0\!\!\! &{{\alpha _{13}}}\!\!\! &{{\alpha _{14}}}\!\!\! &{ - 3{\alpha _{14}}}\!\!\! &{ - 3{\alpha _{13}}} \\
  0&\!\!\! 0\!\!\! &{{\alpha _{14}}}\!\!\! &{ - {\alpha _{13}}}\!\!\! &{3{\alpha _{13}}}\!\!\! &{ - 3{\alpha _{14}}}
\end{array}}\!\!\! \right)\!\!\!\!
\left(\!\!\! {\begin{array}{*{20}{c}}
  {{{({{\tilde k}_{e\!f\!f}})}_{X\!Z\!Z\!Z}}} \\
  {{{({{\tilde k}_{e\!f\!f}})}_{Y\!Z\!Z\!Z}}} \\
  {{{({{\tilde k}_{e\!f\!f}})}_{X\!X\!X\!Z}}} \\
  {{{({{\tilde k}_{e\!f\!f}})}_{Y\!Y\!Y\!Z}}} \\
  {{{({{\tilde k}_{e\!f\!f}})}_{X\!X\!Y\!Z}}} \\
  {{{({{\tilde k}_{e\!f\!f}})}_{X\!Y\!Y\!Z}}}
\end{array}} \!\!\!\right) \nonumber
\\
\end{eqnarray}
and
\begin{widetext}
\begin{eqnarray}
 &&\left( {\begin{array}{*{20}{c}}
  {{c_0}} \\
  {{c_2}} \\
  {{s_2}} \\
  {{c_4}} \\
  {{s_4}}
\end{array}} \right) = \left( {\begin{array}{*{20}{c}} \setlength{\arraycolsep}{0.1pt}
  {{\beta _1}}&{{\beta _1}}&{2{\beta _1}}&{{\beta _2}}&{{\beta _2}}&0&0&0 \\
  {{\alpha _3}}&{ - {\alpha _3}}&0&{{\alpha _4}}&{ - {\alpha _4}}&{{\alpha _5}}&{{\alpha _5}}&{{\alpha _6}} \\
  { - \frac{1}{2}{\alpha _5}}&{\frac{1}{2}{\alpha _5}}&0&{ - \frac{1}{2}{\alpha _6}}&{\frac{1}{2}{\alpha _6}}&{2{\alpha _3}}&{2{\alpha _3}}&{2{\alpha _4}} \\
  { - {\alpha _7}}&{ - {\alpha _7}}&{6{\alpha _7}}&0&0&{{\alpha _8}}&{ - {\alpha _8}}&0 \\
  { - \frac{1}{4}{\alpha _8}}&{ - \frac{1}{4}{\alpha _8}}&{\frac{3}{2}{\alpha _8}}&0&0&{ - 4{\alpha _7}}&{4{\alpha _7}}&0
\end{array}} \right)\left( {\begin{array}{*{20}{c}}
  {{{({{\tilde k}_{eff}})}_{XXXX}}} \\
  {{{({{\tilde k}_{eff}})}_{YYYY}}} \\
  {{{({{\tilde k}_{eff}})}_{XXYY}}} \\
  {{{({{\tilde k}_{eff}})}_{XXZZ}}} \\
  {{{({{\tilde k}_{eff}})}_{YYZZ}}} \\
  {{{({{\tilde k}_{eff}})}_{XXXY}}} \\
  {{{({{\tilde k}_{eff}})}_{XYYY}}} \\
  {{{({{\tilde k}_{eff}})}_{XYZZ}}}
\end{array}} \right) \label{Eq:23}
\end{eqnarray}
\end{widetext}
respectively. In the following analysis, we use Eqs. (17) and (18) to search for LIV.

\section{Lorentz invariance violating torque in torsion pendulum experiments}

Most laboratory gravity experiments employ a low-frequency torsion pendulum. Fig.1 shows a simple I-shaped torsion pendulum consisting of three rectangular glass blocks. The fiber is under tension from the weight of the pendulum, which is, in turn, free to rotate about the axis of the fiber. Specialized test masses, in the form of thin plates with horizontal stripes of alternating density, are attached to the vertical sides of the blocks (e.g., $\rm{Wt_1}$ in Fig. 1). If a force acts on one test mass (generated, for example, by another striped plate $\rm{Ws_1}$ brought into close proximity), the fiber will twist. The angular deflection reveals the interaction strength. The restoration torsion constant of a thin fiber can be quite small, typically on the order of $ 10^{-9}$ $\rm{ Nm/rad}$.

We now consider the LIV interaction between the test mass $\rm{Wt_1}$ and source mass $\rm{Ws_1}$, with density ${\rho _1}$ and ${\rho _2}$ respectively. In the SME laboratory frame, the origin is at the center of the pendulum. Supposing the test mass element $d{m_1} = {\rho _1}d{V_1}$ to be located at position ($x_1$,$ y_1$,$ z_1$) with ${x_1} = {L_1}\cos\theta$ and ${y_1} = {L_1}\sin\theta$ and taking the derivative with respect to $\theta$, we can obtain the LIV torque on the torsion pendulum as
\begin{eqnarray}\label{equation24:eps}
  {\tau _{LV}} \!\!=&& \!\!\!\!\!\!G{\rho _1}{\rho _2}\!\!\iint {d{V_1}d{V_2}}\frac{\partial }{{\partial \theta }}\frac{{\bar k(\hat r,T)}}{{{r^3}}} \hfill  \\
  {\text{     }} =&&\!\!\!\!\!\! G{\rho _1}{\rho _2}\!\!\iint\!\! {d{V_1}d{V_2}}\!\!\left[ {{x_1}\frac{\partial }{{\partial {y_1}}}\frac{{\bar k(\hat r,T)}}{{{r^3}}} - {y_1}\frac{\partial }{{\partial {x_1}}}\frac{{\bar k(\hat r,T)}}{{{r^3}}}} \right]\nonumber
\end{eqnarray}
with the source mass element $d{m_2} = {\rho _2}d{V_2}$ located at position ($x_2$,$ y_2$,$ z_2$). The six-dimensional integration is performed over the volumes for test mass $m_1$ and source mass $ m_2$.

\begin{figure}
\includegraphics[width=0.38\textwidth]{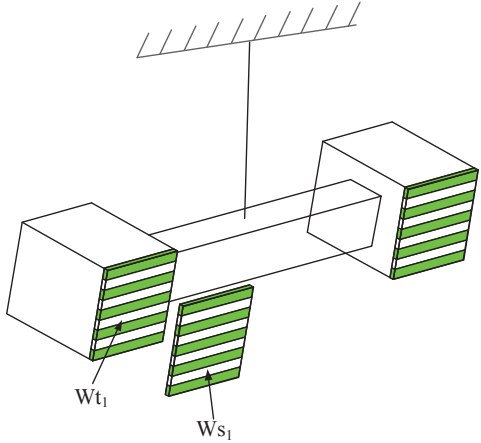}
\caption{\label{fig:1}(color online) The simple I-shape torsion pendulum. The source mass $\rm{Ws_1}$ produces an LIV force on test mass $\rm{Wt_1}$, resulting in a torque on the fiber.}
\end{figure}

We introduce the following 14 transfer coefficients $\Lambda_j$ for $j = 1,2, \cdots ,14$;
\begin{equation}
{\Lambda _j} = G{\rho _1}{\rho _2}\!\!\iint\frac{\partial }{{\partial \theta }}\frac{{{\alpha _j}(\hat r,\chi )}}{r^3}{d{V_1}d{V_2}}.\label{Eq:25}
\end{equation}
Using these coefficients in place of the $\alpha_{j}$ in Eqs. (17) and (18), we may denote the LIV torque $\tau _{LV}$ on the pendulum by
\begin{equation}
{\tau _{LV}}\!\! = {C_0}\!\! +\!\! \sum\limits_{m = 1}^4 {[{C_m}\cos (m{\omega _ \oplus }T) \!+\! {S_m}\sin (m{\omega _ \oplus }T)]}. \label{Eq:26}
\end{equation}
The nine Fourier amplitudes of torque are
 \begin{widetext}
\begin{eqnarray}\label{equation27:eps}
\left( {\begin{array}{*{20}{c}}
{{C_0}}\\
{{C_2}}\\
{{S_2}}\\
{{C_4}}\\
{{S_4}}
\end{array}} \right) = \left( {\begin{array}{*{20}{c}}
{{\Lambda _1}}&{{\Lambda _1}}&{2{\Lambda _1}}&{{\Lambda _2}}&{{\Lambda _2}}&0&0&0\\
{{\Lambda _3}}&{ - {\Lambda _3}}&0&{{\Lambda _4}}&{ - {\Lambda _4}}&{{\Lambda _5}}&{{\Lambda _5}}&{{\Lambda _6}}\\
{ - \frac{1}{2}{\Lambda _5}}&{\frac{1}{2}{\Lambda _5}}&0&{ - \frac{1}{2}{\Lambda _6}}&{\frac{1}{2}{\Lambda _6}}&{2{\Lambda _3}}&{2{\Lambda _3}}&{2{\Lambda _4}}\\
{ - {\Lambda _7}}&{ - {\Lambda _7}}&{6{\Lambda _7}}&0&0&{{\Lambda _8}}&{ - {\Lambda _8}}&0\\
{ - \frac{1}{4}{\Lambda _8}}&{ - \frac{1}{4}{\Lambda _8}}&{\frac{3}{2}{\Lambda _8}}&0&0&{ - 4{\Lambda _7}}&{4{\Lambda _7}}&0
\end{array}} \right)
\left( {\begin{array}{*{20}{c}}
  {{{({{\bar k}_{eff}})}_{XXXX}} - {{({{\bar k}_{eff}})}_{ZZZZ}}} \\
  {{{({{\bar k}_{eff}})}_{YYYY}} - {{({{\bar k}_{eff}})}_{ZZZZ}}} \\
  {{{({{\bar k}_{eff}})}_{XXYY}} - \frac{1}{3}{{({{\bar k}_{eff}})}_{ZZZZ}}} \\
  {{{({{\bar k}_{eff}})}_{XXZZ}} - \frac{1}{3}{{({{\bar k}_{eff}})}_{ZZZZ}}} \\
  {{{({{\bar k}_{eff}})}_{YYZZ}} - \frac{1}{3}{{({{\bar k}_{eff}})}_{ZZZZ}}} \\
  {{{({{\bar k}_{eff}})}_{XXXY}}} \\
  {{{({{\bar k}_{eff}})}_{XYYY}}} \\
  {{{({{\bar k}_{eff}})}_{XYZZ}}}
\end{array}} \right)
\end{eqnarray}
for even harmonics and
\begin{eqnarray} \label{Eq:28}
\left( {\begin{array}{*{20}{c}}
  {{C_1}} \\
  {{S_1}} \\
  {{C_3}} \\
  {{S_3}}
\end{array}} \right) = \left( {\begin{array}{*{20}{c}}
  {{\Lambda _9}}&{{\Lambda _{10}}}&{{\Lambda _{11}}}&{{\Lambda _{12}}}&{{\Lambda _{12}}}&{{\Lambda _{11}}} \\
  { - {\Lambda _{10}}}&{{\Lambda _9}}&{ - {\Lambda _{12}}}&{{\Lambda _{11}}}&{{\Lambda _{11}}}&{ - {\Lambda _{12}}} \\
  0&0&{{\Lambda _{13}}}&{{\Lambda _{14}}}&{ - 3{\Lambda _{14}}}&{ - 3{\Lambda _{13}}} \\
  0&0&{{\Lambda _{14}}}&{ - {\Lambda _{13}}}&{3{\Lambda _{13}}}&{ - 3{\Lambda _{14}}}
\end{array}} \right)
\left( {\begin{array}{*{20}{c}}
  {{{({{\bar k}_{eff}})}_{XZZZ}}} \\
  {{{({{\bar k}_{eff}})}_{YZZZ}}} \\
  {{{({{\bar k}_{eff}})}_{XXXZ}}} \\
  {{{({{\bar k}_{eff}})}_{YYYZ}}} \\
  {{{({{\bar k}_{eff}})}_{XXYZ}}} \\
  {{{({{\bar k}_{eff}})}_{XYYZ}}}
\end{array}} \right)
\end{eqnarray}
for odd harmonics. Any experimental design for searching LIV should make ${\Lambda _j}$ as large as possible.
\end{widetext}

\section{Experimental design with periodic striped geometry}

The most effective geometry for detecting Yukawa-type non-Newtonian gravity is plane-on-plane. This allows one to place the most matter within a given minimum separation between the test and source masses, and makes it practical to place a conducting membrane between the test and source masses for an electrostatic shield. However, the plane-on-plane configuration is not optimal for testing LIV. The LIV force is suppressed for parallel plate geometry, and vanishes altogether in the limit where one plate has infinite area \cite{9}. A more effective geometry for testing LIV is realized with the striped plates as shown in Fig. 1. Here we assume that the periodic density variations are along the $z$ direction.

The inverse-cubic behavior of the potential in Eq. (1) leads to an inverse-quartic force between two point masses. The rapid growth of the force at small distances suggests that the best sensitivities to LIV can be achieved in short-range experiments, which measure the deviation from the Newton gravitational force between two masses. On the other hand, gravitational signals at sufficiently short distances are expected to be overwhelmed by electrostatic, acoustic and Casimir force backgrounds. In the HUST 2011 short-range experiment, which explored the distance range between 0.4 mm and 1.0 mm, these effects were controlled to below the level of the Newtonian background and instrumental thermal noise. Therefore, we base our design for searching the LIV effect on the HUST 2011 experiment, which, in addition, is especially suited to further reduction of the Newtonian background.

\subsection{Test masses and torsion pendulum}

Generally, a high sensitivity test of LIV at millimeter ranges requires that the thickness of the test and source masses, the gap between them , and the source mass amplitude be on the same order of millimeters. The shape of the masses should adopt the striped geometry as shown in Fig. 2 to increase the LIV signal. The width of each strip should also be of the same order of millimeters. The general design is based on the I-shaped pendulum used in our previous work \cite{15,16}. The pendulum, shown in Fig. 2, consists of three rectangular glass blocks. Two blocks, each $23 \times 19.8 \times 19.8$ $\rm{mm}$, support the test masses at each end, and are joined by the third block measuring $64 \times 11 \times 18$ $\rm{mm}$. Ten pure tungsten strips, each measuring  $1.3 \times 19.8 \times 2.2$ $\rm{mm}$, are attached to the vertical surface of each end block in a parallel arrangement with equal spacing, to serve as test masses ($\rm{Wt_1}$ and$ \rm{Wt_2}$). To keep the surface of the test masses flat, glass strips of thickness $1.3 \times 19.8 \times 2.2$ $\rm{mm}$, are inset in the gaps between the tungsten strips. The pendulum; with a total mass of about 70 $\rm{g}$ is suspended by a 25 $\rm{\mu m}$ diameter tungsten fiber.

\begin{figure}[tbp]
\includegraphics[width=0.450\textwidth]{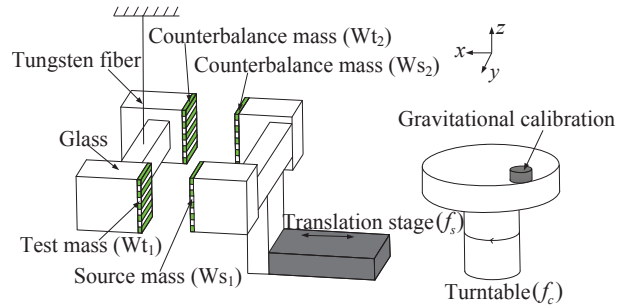}
\caption{\label{fig:2}(color online) Schematic drawing of the experimental design. All the test and source masses are designed with periodic striped geometry in the $z$ direction. The separation between the tungsten test and source masses is modulated with a translation stage with frequency $f_s$, and the sensitivity of the pendulum is calibrated gravitationally with a copper cylinder rotated nearby at frequency $f_c$.}
\end{figure}

\subsection{Source masses and translation stage}

The source mass platform, facing to the pendulum, is also designed with the same I-shaped structure. Two source masses ($\rm{Ws_1}$ and $ \rm{Ws_2}$) with the same sizes as the test masses ($\rm{Wt_1} $ and $ \rm{Wt_2}$), are adhered to the two end glass blocks with dimensions $19.8 \times 19.8 \times 19.8$ $\rm{mm}$. In our design, the source masses are opposite to the test masses, and all of them have the same shape. Therefore, there are two sets of the face-to-face structure for the masses, which can be denoted by the left set ($\rm{Wt_1}$, $\rm{Ws_1}$) and the right set ($\rm{Wt_2}$, $\rm{Ws_2}$), respectively. In this design, we attempt to increase the LIV signal and meanwhile decrease the Newton force signal. After a comprehensive consideration, we came up with the following design: in the both two sets, the positions of the test masses ($\rm{Wt_1}$ and $\rm{Wt_2}$) are kept invariant, and the source mass ($\rm{Ws_1}$) in the left set is shifted up for $a$ of the width of the strip (shifting for a phase $A$), while the source mass ($\rm{Ws_2}$) in right set is shifted down for $b$ of the width of the strip (shifting for a phase $B$). Note that when the total shift for the source masses is kept $a+b\!=\!1$ ($|A|+|B|\!=\!\pi$), a doubled LIV torque on the pendulum can be produced, since the LIV force acting on $\rm{Wt_2}$ due to $\rm{ Ws_2}$ is approximately in the opposite direction to the force acting on $ \rm{Wt_1}$ due to $\rm{Ws_1}$. However, to compensate the Newtonian force well, we finally make a symmetric design, choosing $|A|\!=\!|B|\!=\!\pi/2$, i.e. $ \rm{Ws_1}$ is shifted up half of the width of the strip (1.1 mm in positive z-axis, or shifting $+ \pi /2$) as shown in Fig.3, and $\rm{Ws_2}$ is shifted down half of the width of the strip (1.1 mm in negative z-axis, or shifting $-\pi /2$) as shown in Fig.4. In this design, the Newtonian force between $ \rm{Wt_1}$ and $\rm{ Ws_1}$ was strictly counteracted by interaction between $\rm{Wt_2}$ and $\rm{Ws_2}$. This"anti-symmetric" design strongly suppresses the Newtonian gravitational interaction, while the striped geometry enhances the signal of LIV .

The beam which supports the source and counter mass is symmetrically supported by a glass block, which is mounted on a motor-driven translation stage. The measured signal in the experiment is usually the variation of the torque as the gap between test mass and source mass changes. The experiment will monitor the interaction between test mass and source mass for separations ranging from 0.4 to 1.0 mm.

\begin{figure}[tbp]
\includegraphics[width=0.37\textwidth]{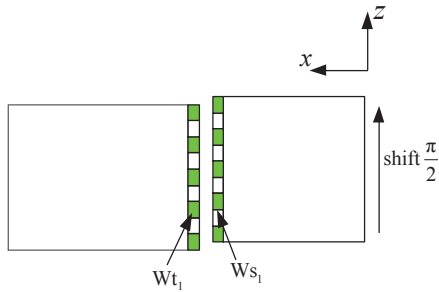}
\caption{\label{fig:3}(color online) Relative positions for test mass $ \rm{Wt_1}$ and source mass $\rm{Ws_1}$ in $x-z$ plane. The position of $\rm{ Ws_1}$ is shifted up half of the width of the strip ( $+ \pi /2$ in phase).}
\end{figure}

\begin{figure}[tbp]
\includegraphics[width=0.36\textwidth]{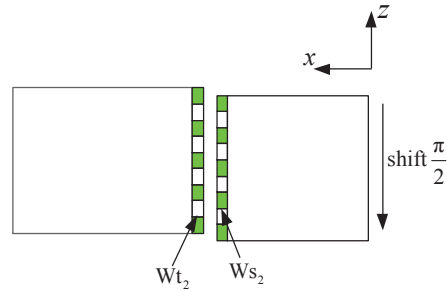}
\caption{\label{fig:4}(color online) Relative positions for test mass $\rm{ Wt_2}$ and source mass $\rm{Ws_2}$ in $x-z$ plane. The position of $\rm{Ws_2}$ is shifted down half of the width of the strip ( $- \pi /2$ in phase).}
\end{figure}

\subsection{Expected signal}

The experimental design leads to a null test of the expected LIV signal. In the experiment, the average gap is set to 0.7 mm, and the motor translation stage is operated continuously with amplitude 0.3 mm at frequency $f_s$. Meanwhile, a copper cylinder mounted on an external turntable is rotated synchronously for gravitational calibration at  frequency $f_c $. The motion of the pendulum is controlled by using a proportional-integral-differential feedback system. In this case, the tungsten fiber is always untwisted during the measurement; the feedback voltage reflects the changes of all effective torques experienced by the pendulum. The measured LIV torque would be
\begin{equation}
\tau _{measured}^z(T) = {\tau _{LV}}(T)\cos (2\pi {f_s}T + \varphi ),\label{Eq:29}
\end{equation}
where $\varphi$ is the initial phase of the separation modulation, which can be determined according to the operation procedure in the experiment. It should be noted that the modulation frequency $ f_s $ can be set far larger than sidereal frequency, such as ${f_s} = 1/500s$. Then in data processing, $\tau _{LV}(T)$ can be taken to be constant in each modulation period $ T_s$.

Using Eqs. (26)-(28), $\tau _{LV}(T)$ can be further expressed as a Fourier series in the sidereal time $T$, and the relations between the nine Fourier amplitudes and the ${{({\bar k_{eff}})_{JKLM}}}$ are described by the 14 coefficients $\Lambda _j$ for $j = 1,2, \cdots ,14$ in Eq. (25). Although the calculation of $\Lambda _j$ between two rectangular plates needs a 6-dimensional integral, the 3-dimension integral for one plate and a single point can be carried out analytically. The result is derived in Appendix A. Calculation of the remaining 3D integral for  $\Lambda _j$ between two finite rectangular plates is carried out numerically. We perform the integration over the complete experimental design in Fig. 2, including the glass blocks, to obtain values for each $\Lambda _j$ with $j=1,2,\cdots,14$. Fig. 5 shows results for the largest $\Lambda _j$ for each of the signal harmonics (including DC), plotted as a function of the gap between the test masses and source masses. In the plot, the  $\Lambda _j$ are shown relative to their value at the assumed minimum gap (0.4 mm). We note that $\Lambda _2$, $\Lambda _4$, $\Lambda _7$ are the largest terms for the even harmonics, and $\Lambda _{11}$, $\Lambda _{13}$ are the largest for the odd harmonics.
\begin{figure}[!h]
\includegraphics[width=0.48\textwidth]{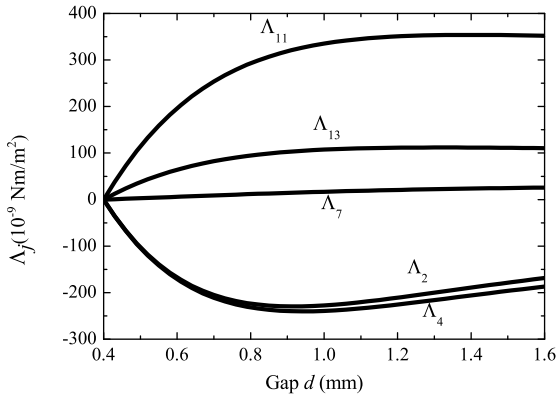}
\caption{\label{fig:5} Curves of $\Lambda _j(d)$ for $j=2, 4, 7, 11, 13$, representing the major terms for even and odd harmonics.}
\end{figure}

The sinusoidal modulation of the  source mass will result in an approximate amplitude of:
\begin{equation}
\Delta {\Lambda _j} \approx \frac{1}{2}[{\Lambda _j}({d_{\min }}) - {\Lambda _j}({d_{\max }})].\label{Eq:30}
\end{equation}
The total expected LIV torque acting on the closed-loop pendulum is
\begin{widetext}
\begin{eqnarray}\label{equation31:eps}
{\tau _{LV}}(T) \!\!= \!\!\!\!&&\sum\limits_{j = 1}^2 {\Delta {\Lambda _j}{{\bar k}_j}}  + \sum\limits_{j = 3}^6 {\Delta {\Lambda _j}{{\bar k}_j}} \cos (2{\omega _ \oplus }T)
 + \sum\limits_{j = 7}^8 {\Delta {\Lambda _j}{{\bar k}_j}} \cos (4{\omega _ \oplus }T)+\sum\limits_{j = 9}^{12} {\Delta {\Lambda _j}{{\bar k}_j}} \cos ({\omega _ \oplus }T)
 + \sum\limits_{j = 13}^{14} {\Delta {\Lambda _j}{{\bar k}_j}} \cos (3{\omega _ \oplus }T)\nonumber\\
 &&{\rm{}}+( - \frac{1}{2}\Delta {\Lambda _5}{{\bar k}_3} - \frac{1}{2}\Delta {\Lambda _6}{{\bar k}_4} + 2\Delta {\Lambda _3}{{\bar k}_5}
+ 2\Delta {\Lambda _4}{{\bar k}_6})\sin (2{\omega _ \oplus }T)+ (\frac{1}{4}\Delta {\Lambda _8}{{\bar k}_7}
  - 4\Delta {\Lambda _7}{{\bar k}_8})\sin (4{\omega _ \oplus }T){\rm{}}+( - \Delta {\Lambda _{10}}{{\bar k}_9} \nonumber \\
 &&+ \Delta {\Lambda _9}{{\bar k}_{10}} - \Delta {\Lambda _{12}}{{\bar k}_{11}} + \Delta {\Lambda _{11}}{{\bar k}_{12}})\sin ({\omega _ \oplus }T)
+ (\Delta {\Lambda _{14}}{{\bar k}_{13}} - \Delta {\Lambda _{13}}{{\bar k}_{14}})\sin (3{\omega _ \oplus }T)\nonumber
\\
\end{eqnarray}
\end{widetext}
with amplitudes $\Delta {\Lambda _j}$ listed in third column of Table I. According to FIG. 5, the transfer coefficients $\Lambda_{j}$ vary with the gap $d$ between the test mass and source mass, especially changing quickly with decreasing $d$ when smaller than 0.8 mm. The experiment is undertaken by implementing a dynamic modulated measurement of the gap distance, thus the value of $d$ is designed to be in the sensitive range, between 0.4 mm to 1.0 mm, which is also the order of the strip width (1.3mm).

\begin{table}[!t]
\caption{\label{tab:pg_I} Transfer coefficients $\left| {\Delta {\Lambda _j}} \right|$ for the LIV in the design with periodic strips in $z$ axis direction.}
\newcommand{\tabincell}[2]{\begin{tabular}{@{}#1@{}}#2\end{tabular}}
\begin{ruledtabular}
\begin{tabular}{ccccc}
{$m$} &  \tabincell{l}{Transfer \\ coefficients} & \tabincell{c}{This design \\ ($10^{-9}$ $\rm{Nm/m^2}$)} & \tabincell{c} {HUST-2011 \\ ($10^{-9}$ $\rm{Nm/m^2}$)}  & \tabincell{c}{Ratio of\\ $\left| {\Delta {\Lambda _j}} \right|$}\\
  \hline
{0}&$\Delta {\Lambda _1}$ &  17.8&  -1.2 &65 \\
 & $\Delta {\Lambda _2}$ & 109.0 &1.2 & \ \\
  \hline
  2&$\Delta {\Lambda _3}$& 24.2& 1.5 & 19 \\
  &$\Delta {\Lambda _4}$  & 112.7 & -3.8 &  \\
 & $\Delta {\Lambda _5}$ & 2.0 & 0.3 & \\
  &$\Delta {\Lambda _6}$  & 0.4 & 4.7 & \\
  \hline
  4&$\Delta {\Lambda _7}$ & -5.3 & 3.9& 1.4 \\
  &$\Delta {\Lambda _8}$ & -1.4& 0.8 &  \\
  \hline
  1&$\Delta {\Lambda _9}$ & 0.0 & -0.5& 15 \\
  &$\Delta {\Lambda _{10}}$ & -0.9 & -3.3& \\
 & $\Delta {\Lambda _{11}}$ & -139.6 & -8.7 &  \\
  &$\Delta {\Lambda _{12}}$ & -4.5 & -0.2 &  \\
  \hline
 3& $\Delta {\Lambda _{13}}$ &-45.2 & 0.2 &40  \\
  &$\Delta {\Lambda _{14}}$ & -0.9 & 1.1 &  \\
\end{tabular}
\end{ruledtabular}
\end{table}


\begin{table}[!t]
\caption{\label{tab:pg_II} The main errors on the $C_0$ amplitude in the design with periodic strips, which include the metrology errors (absolute tolerance on all sources is taken to be 4 microns) and the statistical error (thermal noise). }
\newcommand{\tabincell}[2]{\begin{tabular}{@{}#1@{}}#2\end{tabular}}
\begin{ruledtabular}
\begin{tabular}{lc}
  Source                                           &\tabincell{c}{ Error in $C_0$ \\(${10^{ - 16}}\rm{ Nm}$)}\\
  \hline
  Thickness of source masses                       &$\rm1.3 $  \\
  Thickness of test masses                       & $\rm1.3 $ \\
  Width of the source masses                       &   $\rm0.3 $  \\
  Width of the test masses                         &  $\rm1.0 $ \\
  Horizontal aligning of source and test system    &   $\rm0.1$  \\
  Height aligning of source and test system        &  $\rm3.1$  \\
  \hline
  Statistical error (thermal noise)                                     &  $ 0.4$ \\
   \hline
  Total                                            & $\rm3.8$  \\
  \end{tabular}
\end{ruledtabular}
\end{table}
For comparison, we also list the value of $\Delta {\Lambda _j}$ in our previous short-range experiment HUST-2011 in the fourth column of Table I, which used planar geometry to search for Yukawa-type non-Newton gravity, with the ratio illustrating the relative suppression of the LIV signal. The transfer coefficients $\Delta {\Lambda _j}$ for the new experimental design have largely improved. As shown in the fifth column of Table I, for the constant term $ C_0$, the transfer coefficients include $\Delta {\Lambda _j}$ with $j = 1,2$. The ratio of $\sqrt {\Delta \Lambda _1^2 + \Delta \Lambda _2^2}$ in this design to that of HUST-2011 is about 65, which means signal for LIV for ${\bar k_1}$ and ${\bar k_2}$ can improve 65 times. For the second harmonic frequency terms $ C_2$ and $ S_2$, the transfer coefficients include $\Delta {\Lambda _j}$ with $j = 3,4,5,6$. The ratio of $\sqrt {\Delta \Lambda _3^2 + \Delta \Lambda _4^2 + \Delta \Lambda _5^2 + \Delta \Lambda _6^2}$ in this design to that of HUST-2011 is about 19, which means signal of LIV for ${\bar k_3}$, ${\bar k_4}$ ,${\bar k_5}$ and ${\bar k_6}$ can improve 19 times. Similarly, the ratio of $\left| {\Delta {\Lambda _j}} \right|$ with $j = 7,8$ for fourth harmonic frequency terms is about 1.4. The ratio of $\left| {\Delta {\Lambda _j}} \right|$ with $j = 9,10,11,12$ is about 15. And the ratio of $\left| {\Delta {\Lambda _j}} \right|$ with $j = 13,14$ is about 40.

\subsection{Error budget}
In searching for the Lorentz-violating signal through the short-range experiments, there are some factors affecting the violating torque, such as the statistical error, the metrology errors in the design and the diurnal fluctuations, which we mainly focus on in the new designed short-range experiment.

For the statistical error, we only analyze thermal noise. Dissipative thermal effects affect the sensitivity of the torque in $ C_0$, $ C_m$ and $ S_m$ ($m = 1,2,3,4$). There are two distinct thermal damping mechanisms, velocity damping and internal damping. Velocity damping is usually caused by gas or eddy-current drag, which results in a white spectrum of torque noise. This can usually be made negligible by operation in a vacuum and low magnetic gradient environment. Internal damping caused by the internal friction of the suspension fiber dominates the noise in a torsion-balance experiment. In this case, the spectral density of thermal noise has a $1/f$ character,
\begin{equation}
 S_{th}^{1/2}(\omega ) = \sqrt {\frac{{4{k_B}{T_B}I\omega _0^2}}{{\omega Q}}} \label{Eq:32}
\end{equation}
where $k_B$ is Boltzmann's constant, ${\omega _0}$ is the free resonance frequency, $T_B$ is the temperature, $Q$ is the quality factor and $I$ is the rotation inertia of the pendulum. If the detection bandwidth of the detective torque is  $\Delta f \approx {t^{ - 1}}$, the reciprocal of the measure time interval $ t$, the minimum detectable value of the torque at frequency ${\omega _ \oplus }$,
\begin{equation}
{\tau _{\min }}({\omega _ \oplus }) = \sqrt {\frac{{4{k_B}TI\omega _0^2}}{{t{\omega _ \oplus }Q}}}.\label{Eq:33}
\end{equation}
Taking integration time $t \approx 10$ days, $I{\omega _0}^2 \approx 1 \times {10^{ - 8}}$ $\rm{Nm/rad}$, $Q \approx 1500$, we get ${\tau _{\rm{min} }} \approx 0.4 \times {10^{ - 16}}$ $\rm{Nm}$ at room temperature, this represents the uncertainty of $ C_0$, $ C_m$ and $S_m$ introduced by thermal noise.

Although the ``anti-symmetric" design based on opposite phases can effectively eliminate the Newtonian torque, there are potentially many errors in the machining and aligning of the masses. In general, the pendulum and source masses can be machined with dimensional tolerances on the order of microns, though errors even at this level can produce non-negligible Newtonian effects. However, Newtonian torque is independent of sidereal time $T$ and only affects the constant term $ {C_0}$. Table II lists the main metrology uncertainties used in the numerical model, which translate into the uncertainty of $C_0$. The largest uncertainty comes from the alignment of the height of the source mass system relative to the height of the torsion pendulum. We assume that the position alignment accuracy along both horizontal and vertical directions is 4 microns. Other errors, such as test mass and source mass width and thickness, are also important. The total metrology error is estimated to be $3.8 \times {10^{ - 16}}$ $\rm{Nm}$. Thus, compared with the statistical error (thermal noise), the uncertainty of ${C_0}$ is dominated by the systematic error.

In the short-range experiments, the diurnal fluctuations mainly include the temperature, pressure and electrostatic background fluctuations. These fluctuations affect the torque mainly in two ways: 1) they changes the dimensions and relative positions of the pendulum and the attractor, which lead to a variation of the amplitude of the $f_{s}$ Newtonian torque, consequently contributing the uncertainties to the torque amplitude modes $C_{m}$ and $S_{m}$. Usually, these influences are so small, which will be discussed simply in the following paragraphs. 2) they change the equilibrium position of the torsion balance, leading directly to a torque variation through the corresponding temperature/pressure/electrostatic-to-torque coefficient. Since we focus on the sidereal Lorentz variation signal, which is modulated to the frequency $f_{s}$, the relevant concern therefore is the diurnal variation of the amplitude of the $f_{s}$ torque signal and its harmonics.
Based on the typically related data monitors in our laboratory, we can give a rough upper limit to the errors of the torque amplitude modes $C_{m}$ and $S_{m}$ after extracting the temperature/pressure/electrostatic amplitude fluctuation of the $f_{s}$ torque signal. We will analyze the corresponding systematic errors respectively as below:

For the temperature fluctuation, the largest variation of the geometric parameters due to the temperature fluctuation is the relative height between the pendulum and the attractor, which can be taken to be $0.04$ $\mu m$ for the semidiurnal variation \cite {17}, resulting in a variation of $ < \!0.03 \!\times\! {10^{ - 16}}$ ${\rm{Nm}}$ in the amplitude of the $f_{s}$ Newtonian torque; for typical parameter values, the amplitude fluctuation of the $f_{s}$ temperature signal can be assumed as $3$ $\mu K$, resulting in a $f_{s}$ torque noise of  $ <\! 0.07 \!\times\! {10^{ - 16}}$ ${\rm{Nm}}$ at the $2\sigma $ level with the temperature-to-torque coefficient of $1.1 \!\times\! {10^{ - 12}}$ Nm/K.

For the pressure and electrostatic fluctuations, both of them influence the dimensions and relative positions of the pendulum and the attractor slightly, which usually can be negligible. Based on the data in our previous experiments, the amplitude fluctuation of the $f_{s}$ press signal can be taken as 0.004 mbar, resulting in a $f_{s}$ torque noise of $<0.001\!\times\! {10^{-16}}$ $\rm{Nm}$ at the $2\sigma $  level through a pressure-to-torque coefficient of $1.8\!\times\!{10^{-17}} $ Nm/mbar;
For the electrostatic fluctuation, it is usually regarded as a main error sources in short-range experiments \cite{18,19}.
To minimize the electrostatic force between the pendulum and the source masses, we can insert two stretched beryllium-copper membranes between the test masses and source masses to prevent direct electrostatic coupling. The pendulum and the source mass beams can be entirely gold plated and commonly grounded to prevent contact potentials. Furthermore, the average residual differential potential between the pendulum and the two membranes can be measured and then compensated individually in each experimental cycle. At millimeter ranges, the electrostatic background is usually smaller than thermal noise. For typical parameter values, the amplitude fluctuation of the $f_{s}$ electrostatic variation can be reasonably assumed as $2\!\times\!{10^{-7}}$ V \cite{20}, resulting in a $f_{s}$ torque noise of $<\!0.20\!\times\!{10^{-16}} $ Nm at the $2\sigma $  level through a simple calculation.

From the above analysis, compared with the statistical errors (thermal noise about $0.40\times{10^{-16}} $Nm), the systematic errors discussed above are small (see TABLE III). Therefore, thermal noise sets a fundamental limit to the sensitivity of the torque in $ C_m$ and $ S_m$, i.e. the uncertainties for the modes $C_m$ and $S_m$ can be treated as the same size and dominated by statistical errors in experiment, while the uncertainty of ${C_0}$ is dominated by the systematic error.
\begin{table}[!t]
\caption{\label{tab:pg_III} Errors on the $C_m$ and $S_m$ amplitude in the design with periodic strips, which include the diurnal fluctuations and the statistical error (thermal noise).}
\newcommand{\tabincell}[2]{\begin{tabular}{@{}#1@{}}#2\end{tabular}}
\begin{ruledtabular}
\begin{tabular}{lc}
  Source                                           &\tabincell{c}{ Error in $C_m$  and $S_m$ \\(${10^{ - 16}}\rm{ Nm}$)}\\
  \hline
 Temperature fluctuation                       & $<0.07 $  \\
 Pressure fluctuations                       &   $<0.001 $ \\
  Electrostatic fluctuations                  &   $<0.20 $  \\
  \hline
  Statistical error (thermal noise)                            &   $\sim 0.40$ \\
  \hline
  Total                                       & $< 0.45$  \\
  \end{tabular}
\end{ruledtabular}
\end{table}

\section{Another possible design with periodic striped geometry}

the above design with striped geometry has the periodic structure in the $z$ direction. A similar design has the same periodic structure of the source and test masses in the $y$ direction. The design is shown in Fig. 6. As shown in Fig. 7, the position of source Ws1 is shifted to the left of detector $\rm{ Wt_1}$ by half the width of a strip; the position of source $\rm{Ws_2}$ is shifted to the right of detector $\rm{Wt_2}$ by the same amount. Compared to be LIV force acting on $\rm{Wt_1}$ due to $\rm{ Ws_1}$, the force acting on $\rm{ Wt_2}$ due to $\rm{Ws_2}$ is approximately in the opposite direction, so the total LIV torque on the pendulum is doubled.
\begin{figure}[tbp]
\includegraphics[width=0.450\textwidth]{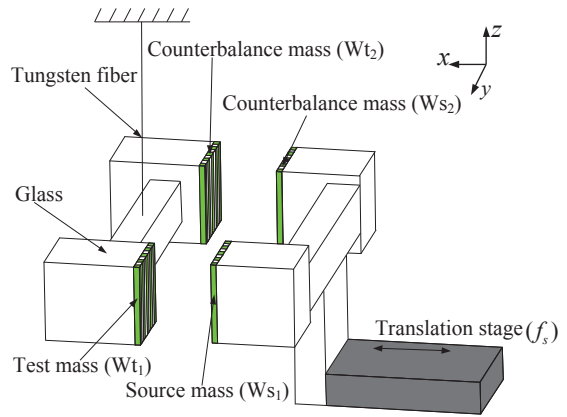}
\caption{\label{fig:6}(color online) Schematic drawing for another possible design. The periodic structure of the test and source masses is in $y$ axis direction.}
\end{figure}

\begin{figure}[tbp]
\includegraphics[width=0.460\textwidth]{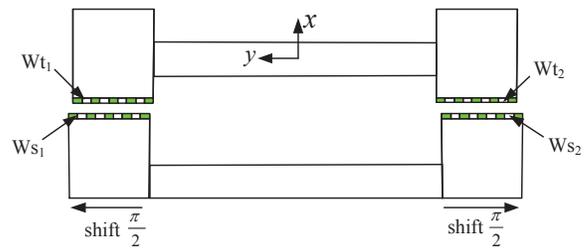}
\caption{\label{fig:7}(color online) Relative positions for test and source masses in $x-y$ plane (top view). The position of $\rm{ Ws_1}$ has shift left half of the width of the strip (shifting $\pi /2$ in phase).  The position of $\rm{ Ws_2}$ has shift right half of the width of the strip (shifting $-\pi /2$ in phase).}
\end{figure}

Assuming all other conditions of this design to be equivalent to the design in Sec. V, the corresponding values of the transfer coefficients $\Lambda$ are shown in Table IV. Only six terms, $\Delta {\Lambda _j}$  with $j = 5,6,8,10,12,14$, are nonzero. This is as expected from Eq. (11), in which only the corresponding terms terms contain odd powers of $y$. Compared with Table I, the LIV signal for the harmonic frequency terms in the design with periodic structure in $y$ axis direction is stronger than that for periodic structure in $z$ axis direction. However, a weakness is that it is insensitive to constant terms $ C_0$, due to the bilateral symmetry of the sources mass.

It is arguable that the design with periodicity in the $z$ direction is perferable (Fig. 2), which is sensitive to all measurable coefficients of $({\bar k_{eff}})_{JKLM}$. On the other hand, one experiment can only provide nine signal components $C_i$ and $S_i$, which are insufficient to independently constrain each of the 14 degrees of freedom in $({\tilde k_{eff}})_{JKLM}$. An additional experiment is required to constrain all $({\tilde k_{eff}})_{JKLM}$ independently; in this sense the design with periodicity in the $y$ direction, would be very useful to improve independent constraints of Lorentz invariance violation.

\begin{table}[!t]
\caption{\label{tab:pg_IV} Transfer coefficients $\Delta {\Lambda _j}$ for the LIV in another design. The striped geometry has the periodic structure in y-axis direction}
\newcommand{\tabincell}[2]{\begin{tabular}{@{}#1@{}}#2\end{tabular}}
\begin{ruledtabular}
\begin{tabular}{cccc}
{$m$} & \tabincell{l}{Transfer \\ coefficients} & \tabincell{c}{This design \\ ($10^{-9}$ $\rm{Nm/m^2}$)}\tabincell{c} & \tabincell{c} {Ratio of $\left| {\Delta {\Lambda _j}} \right|$ \\ to that in HUST-2011 }\\
\hline
{0}&$\Delta {\Lambda _1}$ &0  &0\\
&$\Delta {\Lambda _2}$ &0\\
\hline
{2}&$\Delta {\Lambda _3}$ &0 &67\\
&$\Delta {\Lambda _4}$ &0\\
&$\Delta {\Lambda _5}$ &30.1\\
&$\Delta {\Lambda _6}$ &414.0\\
\hline
 4&$\Delta {\Lambda _7}$ &0 &5.6\\
&$\Delta {\Lambda _8}$ &22.1\\
\hline
 1&$\Delta {\Lambda _9 }$ &0 & 27\\
&$\Delta {\Lambda _{10}}$ &-239.7\\
&$\Delta {\Lambda _{11}}$ &0\\
&$\Delta {\Lambda _{12}}$ &-72.5\\
\hline
 3&$\Delta {\Lambda _{13}}$ &0 &53\\
&$\Delta {\Lambda _{14}}$ &58.9\\
\end{tabular}
\end{ruledtabular}
\end{table}

\section{Summary}

Torsion pendulum experiments that test short-range gravity are sensitive to the Lorentz invariance violations involving quadratic couplings of Riemann curvature. The Lorentz invariance  violation torque includes 14 transfer coefficients $\Delta {\Lambda _j}$ with $j = 1,2 \cdots 14,$ which connect with the 14 measurable coefficients ${{({\bar k_{eff}})_{JKLM}}}$. We decomposed the space $({\bar k_{eff}})_{JKLM}$ (14 dimensions) into 5 subspaces, separating the different harmonics of the Lorentz invariance violation signal and making different harmonic violation signals correspond to different subspaces. From this point of view we optimise our system in the millimeter range, where anticipated backgrounds are small.  After a comprehensive consideration of the form of the LIV force between finite plates, we conclude that a design with striped test masses with periodic density variation will significantly enhance the LIV signal in a torsion pendulum experiment. A geometry with periodic structure in the $z$ direction will have sensitivity to all measurable coefficients of Lorentz violation. An asymmetric design in which the test masses on each side of the pendulum are shifted in opposite directions relative to the source masses, can effectively eliminate the Newtonian gravitational interaction to first order, while greatly enhancing the effects of Lorentz violation. We expect this new design can improve the current constraints on the Lorentz invariance violating coefficients by more than a order of magnitude.

\section{Acknowledgments}

We thank Prof. Joshua C. Long for valuable suggestions and important discussions. This work was supported by the National Natural Science Foundation of China (11275075,11325523 and 91436212) and 111 project (B14030), by the Australian Research CouncilGrant DP160100253.


\section*{Appendix: The analytic expression of Lorentz violation gravitational field for a rectangular plate model}
The calculation of the Lorentz violation force between two rectangular plates needs a 6-dimension integral. The 3-dimensional integral for one plate can be analytically carried out. In this Appendix, we will show that the Lorentz violation gravitational field of a rectangular plate has an analytic expression. In order to simplify the derivation process, we only focus on one term, such as ${{({\bar k_{eff}})_{xxyy}}}$  term in SME lab frame as an example. For the other terms, the derivation can be made in a similar way.

The perturbative potential of a point mass $ m$ is given by
\begin{subequations}
\renewcommand{\theequation}{A\arabic{equation}}
\begin{equation} \label{a1}
{U_{LV,xxyy}}(\vec{r} )\!\! =\!\!   - G\frac{m}{{{r^3}}}\left( {3\!\!  -\!\!  9\frac{{{x^2}\!\!  +\!\!  {y^2}}}{{{r^2}}}\!\!  + \!\! 45\frac{{{x^2}{y^2}}}{{{r^4}}}} \right){({\bar k_{eff}})_{xxyy}}
\end{equation}
at position($x$,$y$,$z$) due to the ${({\bar k_{eff}})_{xxyy}}$ term. To calculate the perturbative potential of a rectangular plate, we perform a 3-dimensional integral over the volume of the plate.

We assume that the density of the rectangular plate is $\rho $ and the dimensions are $2a \times 2b \times 2c$ (in the $x$, $y$ and $z$ directions, respectively) in SME lab frame. We also adopt the plate coordinate system with the origin at the center of the plate. Supposing a unit point mass is at the position ($x$,$y$,$z$) in the plate frame (equal and parallel to the SME lab frame but with different origin), the Lorentz violation gravitational interaction between the rectangular plate and the point mass is given by
\begin{widetext}
\begin{eqnarray}\label{a2}
{U_{LV,xxyy}}(x,y,z) =  - \int_{x - a}^{x + a} {\int_{y - b}^{y + b} {\int_{z - c}^{z + c}}}{{{\frac{{G\rho \left[ {3 - 9\frac{{x_1^2 + y_1^2}}{{x_1^2 + y_1^2 + z_1^2}} + 45\frac{{x_1^2y_1^2}}{{{{\left( {x_1^2 + y_1^2 + z_1^2} \right)}^2}}}} \right]{{({{\bar k}_{eff}})}_{xxyy}}}}{{{{\left( {x_1^2 + y_1^2 + z_1^2} \right)}^{3/2}}}}d{x_1}d{y_1}d{z_1}} } }
\end{eqnarray}
Taking the derivative with respect to $z$, we obtain the Lorentz violation gravitational field in $z$ axis direction
\begin{eqnarray}\label{a3}
{a^z}(a,b,c;x,y,z) \equiv&&\!\!\! {\partial _z}U(x,y,z) \nonumber\\
 =&&\!\!\!  - {({{\bar k}_{eff}})_{xxyy}}G\rho \int_{x - a}^{x + a} {\int_{y - b}^{y + b} {\frac{{\left[ {3 - 9\frac{{x_1^2 + y_1^2}}{{x_1^2 + y_1^2 + {{(z + c)}^2}}} + 45\frac{{x_1^2y_1^2}}{{{{\left[ {x_1^2 + y_1^2 + {{(z + c)}^2}} \right]}^2}}}} \right]}}{{{{\left[ {x_1^2 + y_1^2 + {{(z + c)}^2}} \right]}^{3/2}}}}d{x_1}d{y_1}} } \\
 &&+ {({{\bar k}_{eff}})_{xxyy}}G\rho \int_{x - a}^{x + a} {\int_{y - b}^{y + b} {\frac{{\left[ {3 - 9\frac{{x_1^2 + y_1^2}}{{x_1^2 + y_1^2 + {{(z - c)}^2}}} + 45\frac{{x_1^2y_1^2}}{{{{\left[ {x_1^2 + y_1^2 + {{(z - c)}^2}} \right]}^2}}}} \right]}}{{{{\left[ {x_1^2 + y_1^2 + {{(z - c)}^2}} \right]}^{3/2}}}}d{x_1}d{y_1}} } \nonumber
\end{eqnarray}
The integral can be carried out resulting in an analytical expression. After defining the functions
\begin{eqnarray}\label{a4}
f(x,y,z) \equiv\int_{}^x {\int_{}^y {\frac{1}{{{{({x^2} + {y^2} + {z^2})}^{3/2}}}}dxdy} }{} = \frac{1}{z}\arctan \frac{{xy}}{{z\sqrt {{x^2} + {y^2} + {z^2}} }}
\end{eqnarray}
\begin{eqnarray}\label{a5}
{f_{xx}}(x,y,z) \equiv\!\!\! \int_{}^x \! \! {\int_{}^y \! \! \! \! {\frac{1}{{{{({x^2}\! +\! {y^2} \!+\! {z^2})}^{3/2}}}}\frac{{{x^2}}}{{{x^2} \!+\! {y^2} \!+\! {z^2}}}dxdy} }
{\rm{}} =  \!\!\!\frac{1}{3}\left[ { - \frac{{xy}}{{({x^2} \!+\! {z^2})\sqrt {{x^2} \!+\! {y^2}\! +\! {z^2}} }} \!+ \!\frac{1}{z}\arctan \frac{{xy}}{{z\sqrt {{x^2} \!+\! {y^2} \!+ \!{z^2}} }}} \right]
\end{eqnarray}
\begin{eqnarray}\label{a6}
{f_{yy}}(x,y,z) \!\equiv\!\! \int_{}^x\!\!\! {\int_{}^y \!\!\!\!{\frac{1}{{{{({x^2} \!+ \!{y^2} \!+\! {z^2})}^{3/2}}}}\frac{{{y^2}}}{{{x^2} \!+ \!{y^2} \!+\! {z^2}}}dxdy} }
{\rm{                }} =\!\!\! \frac{1}{3}\left[ { - \frac{{xy}}{{({y^2} \!+\! {z^2})\sqrt {{x^2} \!+\! {y^2}\! +\! {z^2}} }}\! +\! \frac{1}{z}\arctan \frac{{xy}}{{z\sqrt {{x^2} \!+\! {y^2} \!+\! {z^2}} }}} \right]
\end{eqnarray}
\begin{eqnarray}\label{a7}
{f_{xxyy}}(x,y,z) \!\equiv &&\!\! \int_{}^x \!\!\!{\int_{}^y\!\!\! {\frac{1}{{{{({x^2} \!+\! {y^2} \!+\! {z^2})}^{3/2}}}}\frac{{{x^2}{y^2}}}{{{{({x^2}\! +\! {y^2} \!+\! {z^2})}^2}}}dxdy} }\nonumber\\
=&&\!\! \frac{1}{{15}}\left[ { - \frac{{xy[{x^4} \!+\! {{({y^2} \!+ \!{z^2})}^2}\! +\! {x^2}({y^2} \!+\! 2{z^2})]}}{{({x^2} \!+\! {z^2})({y^2}\! + \!{z^2}){{({x^2} \!+\! {y^2}\! +\! {z^2})}^{3/2}}}} \!+\! \frac{1}{z}\arctan \frac{{xy}}{{z\sqrt {{x^2} \!+\! {y^2}\! + \!{z^2}} }}} \right]
\end{eqnarray}
we obtain the analytic expression of the gravitational field in $z$-axis direction
\begin{eqnarray}\label{a8}
{a^z}(a,b,c;x,y,z) = \!\! - G\rho {({\bar k_{eff}})_{xxyy}}F_{xxyy}^z(x,y,z)\left. {} \right|_{x = x - a}^{x = x + a}\left. {} \right|_{y = y - b}^{y = y + b}\left. {} \right|_{z = z - c}^{z = z - c}
\end{eqnarray}
with
\begin{eqnarray} \label{a9}
F_{xxyy}^z(x,y,z)= 3f(x,y,z) - 9{f_{xx}}(x,y,z)- 9{f_{yy}}(x,y,z) + 45{f_{xxyy}}(x,y,z)
\end{eqnarray}
The gravitational field in $x$-axis or $y$-axis direction can be derived in a similar way.
\end{widetext}
\end{subequations}
\end{document}